\documentclass{aa}  
\usepackage{natbib}
\usepackage{graphicx}
\usepackage{txfonts}
\usepackage{lipsum}
\usepackage[colorlinks=true, allcolors=blue]{hyperref}
\usepackage{float}
\usepackage{booktabs}      
\usepackage{adjustbox}     
\usepackage{threeparttable}
\usepackage{placeins}      

\usepackage{lscape}             
\usepackage{booktabs}
\usepackage{placeins}           
                                
\begin{document}
  \title{Astronomy \& Astrophysics \LaTeX\ template}

\title{Evidence of energy conversion in weakly collisional plasma during an interplanetary coronal mass ejection}

\titlerunning{Energy conversion during an ICME}
\authorrunning{Dhamane et al.}

\author{
  Omkar Dhamane\inst{1,2}
  \and Anil Raghav\inst{2}
  \and Simone Benella\inst{1}
  \and Kishor Kumbhar\inst{2}
  \and Raffaella D'Amicis\inst{1}
  \and Oreste Pezzi\inst{3,1}
  \and Utkarsh Sharma\inst{4}
  \and Ashok Silwal \inst{5}
  \and Panini Maurya\inst{6}
  \and Mirko Stumpo\inst{1}
    \and Kalpesh Ghag\inst{2}
  \and Ajay Kumar\inst{2}
  \and Mohit Shah \inst{2,10}
  \and Mariyam Karari\inst{2}
  \and Lynn B. Wilson III\inst{7}
  \and Jia Huang\inst{8}
  \and Daniele Telloni\inst{9}
}

\institute{
  National Institute for Astrophysics, Institute for Space Astrophysics and Planetology, Via del Fosso del Cavaliere 100, I-00133 Roma, Italy
  \and Department of Physics, University of Mumbai, Mumbai, India
  \and Istituto per la Scienza e Tecnologia dei Plasmi, Consiglio Nazionale delle Ricerche (ISTP-CNR), Via G. Amendola 122/D, I-70126 Bari, Italy
  \and Center of Excellence in Space Sciences India, Indian Institute of Science Education and Research Kolkata,
Mohanpur 741246, West Bengal, India
  \and Center for Space Plasma and Aeronomic Research (CSPAR), The University of Alabama in Huntsville, Huntsville, AL 35805, USA
  \and Indian Institute of Geomagnetism (IIG), Kalamboli, New Panvel, Navi Mumbai, 410218, India
  \and NASA Goddard Space Flight Center, Heliophysics Science Division, Greenbelt, MD, USA
  \and Space Sciences Laboratory, University of California, Berkeley, CA, USA
  \and National Institute for Astrophysics, Astrophysical Observatory of Torino, Via Osservatorio 20, I-10025 Pino Torinese, Italy
  \and Department of Space and Climate Physics, Mullard Space Science Laboratory, University College London, Dorking, Surrey RH5 6NT, UK
 }

   \date{Received xx; accepted xx}

\abstract
  {Intervals of enhanced turbulent fluctuations are typically less frequent within the magnetic cloud region of an interplanetary coronal mass ejection (ICME).}
  {We investigate two such intervals inside an ICME observed by the \textit{Wind} spacecraft on 8--9 June 2000 and characterize their associated wave populations.}
  {We focus on spectral analysis and plasma instability analysis, using ion-scale normalized magnetic helicity and polarization properties with respect to the background magnetic field $B_0$.}
  {In the first interval, the ion-scale normalized magnetic helicity shows a left-handed circularly polarized signature. In the second interval, the left-handed signature persists and an additional high-frequency right-handed population appears. The propagation is approximately parallel to $B_0$. The left-handed fluctuations are compatible with Alfv\'en ion-cyclotron (AIC) waves, while the right-handed fluctuations are consistent with fast magnetosonic/whistler (FM/W) waves.}
  {The ICME plasma accesses resonance conditions that support multiple ion-scale wave modes. Evolving anisotropies in the plasma and the approach to marginal stability allow the coexistence of AIC-like and fast-magnetosonic/whistler-like fluctuations, with enhanced electron heating favoring the growth of the FM/W contribution and strengthening the density--magnetic-field magnitude correlation.}


\maketitle



\section{Introduction} \label{sec:intro}

Coronal mass ejections (CMEs) are violent manifestations of solar activity in which large amounts of plasma and magnetic flux are expelled from the corona \citep{webb2012coronal}. Their interplanetary counterparts (ICMEs) \citep{richardson2004identification} often contain magnetic cloud, characterized by enhanced magnetic-field strength, smooth and coherent magnetic-field rotations, and low plasma beta ($\beta \ll 1$). 
The enhanced  magnetic-field strength in magnetic clouds reduces thermal pressure, increases proton and electron gyrofrequencies, and compresses kinetic spatial and temporal scales. At the same time, the ion and electron velocity distribution functions are frequently non-Maxwellian owing to expansion, internal heating, and intermittent reconnection, producing temperature anisotropies that can shift classical thresholds for Alfvén ion–cyclotron (AIC) and fast magnetosonic/whistler (FM/W) instabilities \citep{hellinger2013protons,podesta2011magnetic,woodham2019parallel}.


Contrary to the ambient solar wind, which is continually driven by expansion, turbulence, and stream–stream interactions that can maintain strong temperature anisotropies and enables a broad spectrum of kinetic waves—including AIC, FM/W, mirror, and electron-scale whistler modes (e.g., \citep{he2011oblique,he2011possible,podesta2011magnetic,podesta2013evidence,telloni2019ion}), ICME interiors are unusually smooth structures in which turbulence, velocity shear, and thermal driving are significantly reduced. This maintains the plasma near marginal stability, narrows the available resonance channels, and significantly restricts the wave modes that can grow. Only when local anisotropies exceed instability thresholds, or when electron-temperature variations shift resonance conditions, can ion-kinetic waves be generated. Although observational evidence remains limited, several studies have reported such rare events within ICMEs, including Alfvén ion-cyclotron waves \citep{dhamane2023aic}, torsional and surface Alfvénic modes \citep{raghav2018first,raghav2018does,dhamane2023observation,raghav2023first}, kinetic Alfvén waves \citep{kumbhar2024observation}, and low-frequency fluctuations \citep{siu2015low}. One of the clearest multi-mode demonstrations was provided by \citet{moullard2001whistler}, who observed correlated Langmuir waves, field-aligned whistlers, and electron loss-cone distributions inside a magnetic cloud.

Despite their weak turbulence, magnetic cloud plasmas are not perfectly isotropic \citep{zwickl1982plasma,galvin1987solar}. Variations in magnetic-field strength, expansion-driven cooling, localized heating, and changing particle populations can modify the ion and electron velocity distribution functions, causing  perpendicular emperature ($T_{\perp i}$) and parallel temperature ($T_{\parallel i}$) to evolve differently \citep{farrugia1998possible,dasso2003parametric}. A particularly important source of anisotropy in magnetic clouds arises from their closed magnetic-field topology, which can trap energetic particles and produce bidirectional fluxes that preferentially enhance the parallel temperature \citep{gosling1987bidirectional,marsden1987isee}. Such departures from isotropy supply free energy for the excitation of kinetic instabilities \citep{gary1976proton}, whenever the plasma briefly crosses the relevant stability thresholds. Through wave–particle scattering and energy redistribution, these instabilities regulate the anisotropy itself, establishing a feedback mechanism that maintains the plasma close to marginal stability \citep{shaikh2025temperature}. Because the thresholds for AIC, FM/W, and firehose-like modes are strongly $\beta$-dependent, even small changes in $T_{\perp}/T_{\parallel}$ or in the local plasma environment can activate or suppress different ion-scale wave modes. These considerations naturally connect to the view that ICME plasma may experience transient anisotropies due to expansion, wave activity, or localized shears, which can keep the plasma temporarily unstable and susceptible to kinetic instabilities \citep{shaikh2025temperature}.


The interiors of magnetic clouds are typically quiescent and exhibit reduced levels of magnetic and plasma fluctuations \citep[e.g.,][]{samith1999magnetic,singh2007effects,bhattacharjee2023turbulence}.
 Their coherent flux-rope topology and enhanced magnetic-field strength inhibit the cascade of turbulence to kinetic scales, making instability-driven waves far less common than in typical solar-wind plasma. As a result, intervals of enhanced fluctuation power inside magnetic clouds are rare, yet they offer valuable opportunities to study wave generation and instability dynamics in low-$\beta$, weakly turbulent plasma near marginal stability.

\section{Data and Observation}

We analyzed an ICME event measured by \textit{Wind} spacecraft  \citep{wilson2021quarter} on June 8, 2000.  This event is also available at \textit{Wind} ICME catalog \footnote{\url{https://wind.nasa.gov/ICME_catalog/ICME_catalog_viewer.php}}. We utilized high resolution observations from the Magnetic Field Investigation (MFI) instrument aboard the \textit{Wind} spacecraft \citep{lepping1995wind}, together with plasma measurements from the Three–Dimensional Plasma and Energetic Particle Investigation (3DP) instrument \citep{lin1995three}. We used magnetic field data at 3 s cadence to characterize the large-scale interplanetary conditions, and higher-resolution 92 ms MFI data for spectral and wave polarization analyses. Plasma moments from the 3DP onboard–derived products (3 s resolution) were independently validated using nonlinear Faraday cup fits from the Solar Wind Experiment (SWE; \citealt{ogilvie1995swe}) to ensure that the onboard moments were reliable and not significantly contaminated by instrumental or fitting uncertainties.

\begin{figure*}
	\centering
	\includegraphics[width=16cm]{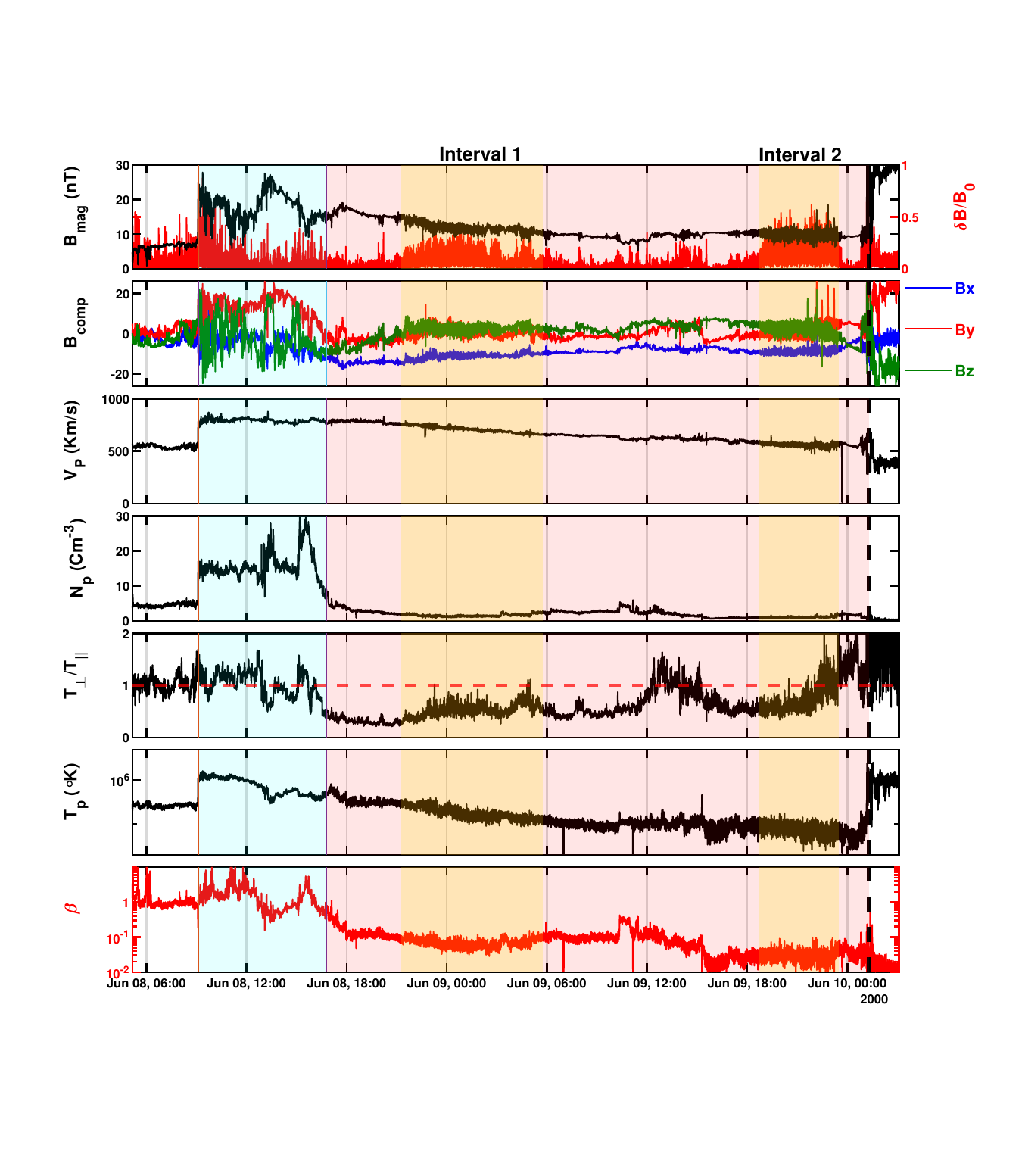}	
	\caption{
Interplanetary parameters associated with the ICME observed on 08–09 June 2000. 
From top to bottom, the panels show: the total magnetic-field magnitude $B_{\mathrm{mag}}$ (in nT, GSE coordinates); the normalized magnetic-field fluctuation amplitude $\delta B/B_{0}$, computed as the centered-difference variation of $\vec{B}$ and normalized by the local background field; the magnetic-field vector components ($B_{\mathrm{comp}}$ in nT)); the proton bulk speed $V_p$ (km s$^{-1}$); proton density $N_p$ (cm$^{-3}$); proton temperature anisotropy $T_{\perp}/T_{\parallel}$; proton temperature $T_{p}$; and plasma beta $\beta$. 
The ICME sheath is shaded in cyan, the magnetic cloud in pink, and the two yellow intervals mark periods of enhanced magnetic variability within the magnetic cloud. The dashed line at the end part indicates the time at which the spacecraft entered the bow shock.
}
 \label{fig:ip2}	
\end{figure*}

The ICME boundaries are taken from the catalog mentioned above.
The temporal variations of the interplanetary magnetic field (IMF) and various plasma parameters are illustrated in Figure \ref{fig:ip2}. The arrival of the ICME shock front is identified by a sharp discontinuity at 09:07 UT on June 8, 2000, marked by abrupt increases in the total magnetic field strength ($B_{\text{mag}}$), proton density ($N_p$), proton temperature ($T_p$), and proton velocity ($V_p$). Following the shock front, a region exhibiting high plasma density and temperature, enhanced magnetic field strength, and significant magnetic field fluctuations is observed. This region corresponds to the ICME sheath and is shaded cyan in Figure \ref{fig:ip2}, consistent with previous observations \citep{richardson2004identification}. Following the shock sheath, the observations indicate a transition into a magnetic cloud-like structure at 16:47 UT on June 8, represented by the pink-shaded region in Figure \ref{fig:ip2}. This region exhibits a gradual decrease in magnetic field strength and solar wind speed, as well as a very low plasma beta ($\beta$), consistent with the characteristics of magnetic clouds \citep{burlaga1982magnetic, zurbuchen2006situ}. 
Furthermore, the smooth decline in velocity and magnetic field strength within the magnetic cloud indicates its expansion, albeit slower than typical ICME events. Determining the precise end boundary of this ICME proves challenging due to the spacecraft crossing the bow shock boundary, which is evident from a sudden rise in the observed magnetic field on 10 June 2000, at 01:25:37 UT, marked by a vertical black dashed line in Figure \ref{fig:ip2}. 

We analyzed 3-second resolution magnetic-field data to identify regions of enhanced magnetic variability within the ICME. Magnetic-field fluctuations were quantified using a centered finite difference,
\[
\delta B_i=\frac{B(t_{i+1}) - B(t_{i-1})}{2},
\]
applied to each component of $\mathbf{B}$. The fluctuation amplitude was then taken as the vector magnitude $|\delta B|$ and normalized by the local background magnetic-field magnitude $B_0$.

The first interval of intense fluctuations spans from 21:15 UT on June 8 to 05:45 UT on June 9, while the second interval extends from 18:40 UT to 23:30 UT on June 9 (shown in yellow). These regions are characterized by enhanced  IMF fluctuations. After the second interval, the fluctuations diminished, and the spacecraft then crossed the Earth's bow shock. These fluctuations are detached from the bow shock, as smooth variations in the magnetic field can be observed between the bow shock and the highly fluctuating second interval.

\section{Results}
\subsection{Spectral Analysis and wave polarization}

\begin{figure*}
        \centering
        \includegraphics[width=\linewidth]{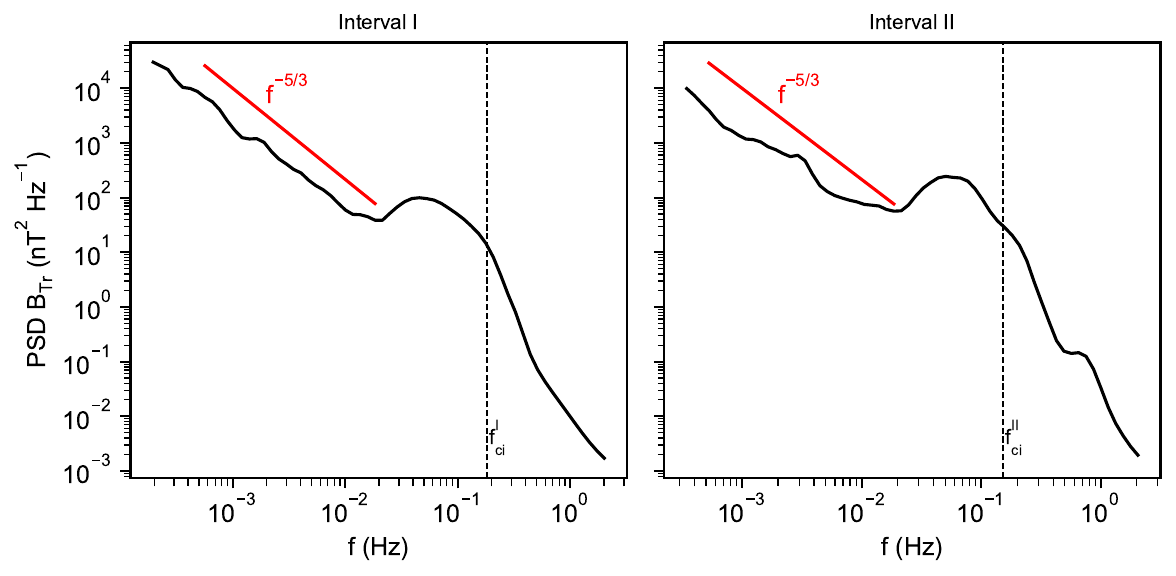}
        \caption{Power spectral density (PSD) of the magnetic field trace for the two intervals shown in Figure~1. The red solid line indicates a reference $-5/3$ slope associated with Kolmogorov-like inertial-range scaling, while the vertical dashed line marks the local proton cyclotron frequency $f_{ci}$.}
        \label{fig:placeholder}
    \end{figure*}
  
To assess the nature of fluctuations, we have computed the power spectral density (PSD) of the trace of the magnetic field fluctuations for both intervals, shown in Figure \ref{fig:placeholder}. At low frequencies, the magnetic-field power spectra are broadly consistent with a Kolmogorov-like $-5/3$ scaling over a limited frequency range. As the frequency approaches ion-kinetic scales ($\sim 10^{-2}$–$10^{-1}$~Hz), the spectra depart from this trend and exhibit a pronounced spectral enhancement. This behavior suggests increased fluctuation power at ion scales, rather than a smooth continuation of the inertial-range cascade into the dissipation range. Such spectral features are commonly associated with the onset of ion-scale physics, where turbulence transitions from MHD-like behavior to a regime influenced by dispersive effects and cyclotron-resonant interactions \citep{bruno2013solar, chen2014ion}. Processes such as kinetic Alfvén waves, proton cyclotron waves, anisotropic ion-scale fluctuations, or localized particle–field interactions may contribute to the redistribution or enhancement of energy at these scales \citep{bruno2014radial, telloni2015radial, telloni2016linking, telloni2020persistence}. Consequently, the spectrum does not steepen monotonically but instead displays a characteristic enhancement near the proton cyclotron frequency ($f_{ci}$). Motivated by this feature, we further investigate whether the enhanced fluctuations are associated with wave activity by examining their Alfvénicity, compressibility, and polarization properties in the following subsections.

The Alfvénic nature of the observed fluctuations was examined through spectral analysis of the Elsässer variables, \( \mathbf{z}^{\pm} \) \citep{elsasser1950hydromagnetic, grappin1991alfvenic, tu1989basic}, defined as \( \mathbf{z}^{\pm} = \mathbf{v} \pm \mathbf{b} \), where \( \mathbf{v} \) is the velocity vector and \( \mathbf{b} \) is the magnetic field vector expressed in Alfvén units. When the background magnetic field is directed toward the Sun, the definition is reversed (\( \mathbf{z}^{\pm} = \mathbf{v} \mp \mathbf{b} \)), so that \( \mathbf{z}^+ \) always indicates outward-propagating Alfvénic modes. The Elsässer variables thus provide an effective framework for visualizing and quantifying Alfvénicity in plasma fluctuations.

To characterize the turbulence, we computed the normalized cross-helicity (\( \sigma_c \)) and the normalized residual energy (\( \sigma_r \)) in the frequency domain similar to \cite{marsch1990radial}:

\[
\sigma_c(f) = \frac{e^{+}(f) - e^{-}(f)}{e^{+}(f) + e^{-}(f)}, \quad 
\sigma_r(f) = \frac{e^{v}(f) - e^{b}(f)}{e^{v}(f) + e^{b}(f)}.
\]

Here, \( e^{\pm}(f) \) are the power spectral densities (PSD) of the Elsässer components, and \( e^{v}(f) \) and \( e^{b}(f) \) denote the kinetic and magnetic power spectral densities, respectively. Both parameters range from \(-1\) to \(+1\). A value of \( \sigma_c = \pm 1 \) corresponds to purely (outward or inward) Alfvénic fluctuations, while intermediate values (\(|\sigma_c| < 1\)) indicate a mixture of counterpropagating or non-Alfvénic modes. Similarly, $\sigma_r = 0$ denotes energy equipartition between kinetic and magnetic components, while $\sigma_r = \pm 1$ corresponds to purely kinetic ($+1$) or purely magnetic ($-1$) fluctuations. For ideal Alfv\'enic modes, $\sigma_C = \pm 1$ and $\sigma_R =0$.

\begin{figure*}[t]
	\centering
		\includegraphics[width = 13cm]{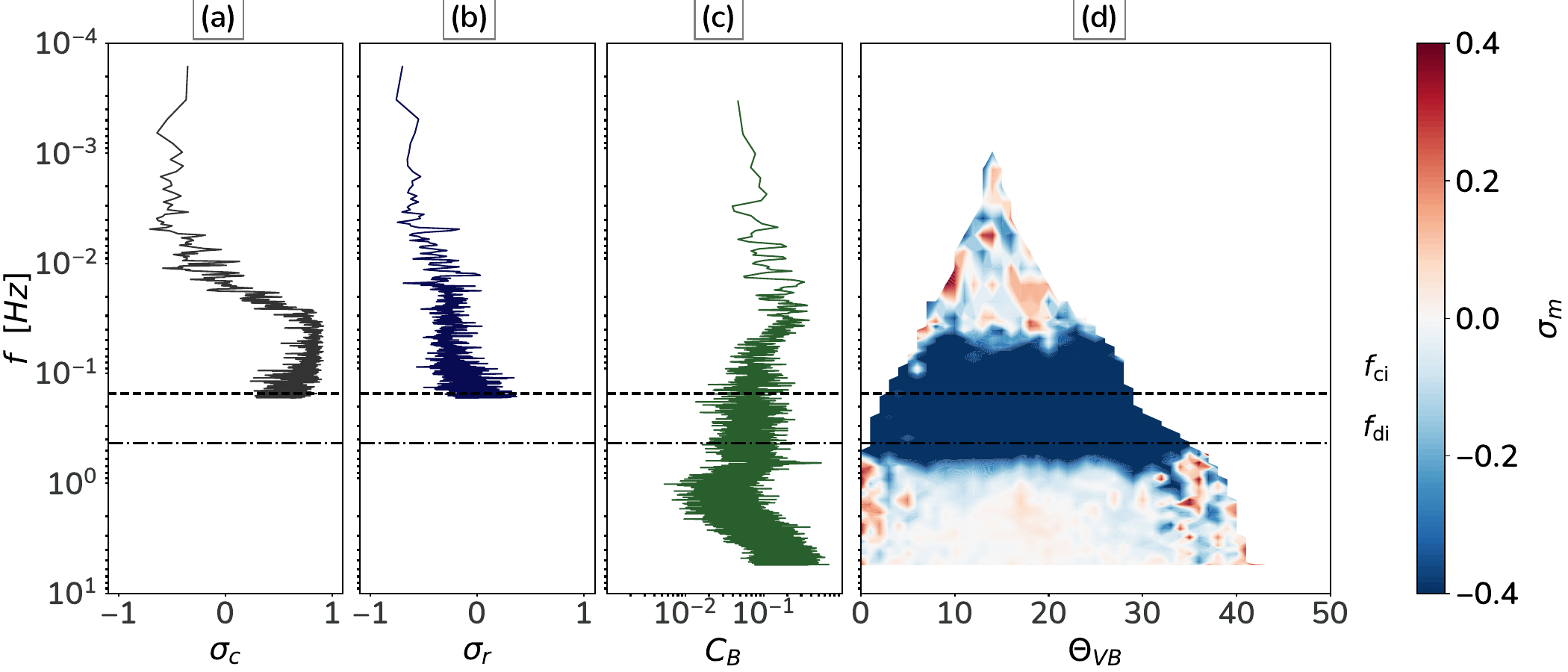}
        \includegraphics[width = 13cm]{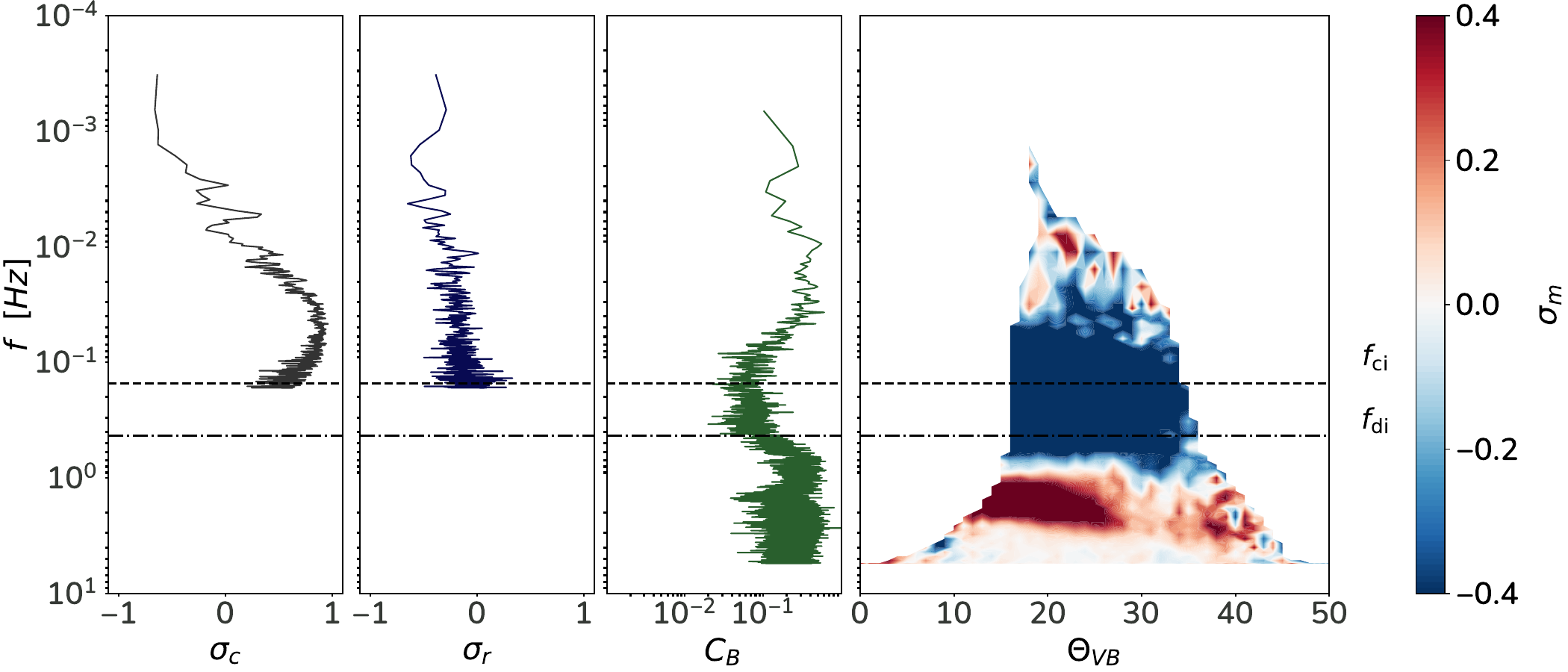}
	\caption{The spectral analysis of (a) $\sigma_c$, (b) $\sigma_r$, (c) magnetic compressibility ($C_B$)  along with  (d) the distribution of $\sigma_m$ as a  function of flow angle $\theta_{VB}$, is shown for the fluctuating interval 1 (top) and 2 (bottom).  For $C_B$ and $\sigma_m$, high-resolution data at 11 Hz are utilized, whereas $\sigma_c$ and $\sigma_r$ were computed using 3-second data. The solid dashed lines represent the ion gyrofrequency ($f_{ci}$) and skin depth ($f_{di}$).
    }
	\label{fig:spectral}
\end{figure*}

Figure~\ref{fig:spectral} depicts the spectral analysis of the highly fluctuating interval highlighted in Figure~\ref{fig:ip2}. Both intervals within the magnetic cloud exhibit similar behavior in the normalized cross helicity, $\sigma_c$, and normalized residual energy, $\sigma_r$. As shown in panel (a), $\sigma_c$ remains close to $-0.5$ at frequencies below \(10^{-2}\)~Hz, indicating a weakly Alfv\'enic state. With increasing frequency, $\sigma_c$ rises sharply and approaches $\sim 0.9$, corresponding to a highly Alfv\'enic regime; a similar frequency-dependent transition is also observed in the second interval. Panel (b) shows that $\sigma_r$ approaches equipartition in the frequency range \(10^{-2}\)–\(10^{-1}\)~Hz, consistent with the more Alfv\'enic fluctuations, with a slight imbalance favoring magnetic energy, suggesting that magnetic fluctuations dominate over kinetic energy at these scales.

Alfvénic turbulence is characterized not only by strong correlations between velocity and magnetic field fluctuations but also by weak magnetic compressibility, implying that fluctuations in the magnetic field magnitude are much smaller than those of the field vector and correspond to nearly spherical polarization \citep{bruno2013solar, matteini2015ion}. In the third vertical panel (c) of Figure~\ref{fig:spectral}, we compute the magnetic field compressibility, $C_{B}$, defined as the ratio of the PSD of the magnetic field magnitude to the trace of the PSD of the magnetic field vector. Weak compressibility is observed during the highly Alfvénic intervals, coincident with the maxima of $\sigma_c$. Due to the availability of high-resolution (11~Hz) magnetic field measurements, it is further evident that beyond ion scales the spectra become slightly less compressible in the first interval, whereas the second interval exhibits a comparatively more compressive behavior.

While this analysis provides important insight into the Alfvénic nature of the fluctuations, the 3~s resolution plasma data do not allow a detailed identification of ion-scale wave modes. To address this limitation, we therefore turn to higher-resolution magnetic field observations and examine the polarization properties of the fluctuations.

To better characterize the ion scale fluctuations, we
use a measure of the magnetic helicity, an invariant of
ideal magnetohydrodynamics (MHD), to analyze the nature of
the magnetic field fluctuations using their polarisation properties. 
The concept of reduced magnetic helicity ($\sigma_m$) introduced by \citet{matthaeus1982measurement} has been instrumental in identifying turbulent fluctuations that resembles wave-like polarization properties in the solar wind. At MHD scales, \(\sigma_m\) approaches zero, while at kinetic scales, it increases toward unity, reflecting strong polarization either left- or right-handed \citep{gary1986low}. However, observed values of \(\sigma_m\) consistently remain below unity, suggesting a mixture of waves or interactions that reduce the normalized magnetic helicity. This reduction hints at complex nonlinear processes involving turbulent interactions among differently oriented kinetic-scale waves. 
This measure can be investigated in the time–frequency domain using the wavelet transform method \citep{torrence1998practical}, which has been subsequently adopted in several studies \citep{he2011possible, he2012reproduction, podesta2011effect, telloni2012wavelet, bruno2015spectral, dhamane2023aic, kumbhar2024observation}.

Using wavelet analysis, we map the angle between the proton bulk velocity and the magnetic field, $\theta_{VB}$, alongside the normalized magnetic helicity, $\sigma_m$, for both intervals. The results are shown in the last vertical panel of Figure~\ref{fig:spectral}. In the first interval (top, panel d), magnetic fluctuations with negative helicity are observed and are associated with propagation angles in the range $0^{\circ} \leq \theta_{VB} \leq 30^{\circ}$. These small propagation angles indicate that this interval is predominantly populated by quasi-parallel ($k_{\parallel}$) fluctuations. Consequently, the positive patches in the helicity spectrum near ion-kinetic scales provide robust evidence for quasi-parallel, left-handed polarized Alfvén/ion-cyclotron (AIC) waves \citep{podesta2011magnetic,he2011oblique,telloni2019ion,dhamane2023aic,fordin2023machine}.

In contrast, the second interval (bottom, panel d) exhibits the simultaneous presence of negative-helicity patches at low frequencies and positive-helicity patches at higher frequencies within a similar range of propagation angles. This represents a clear case of opposite helicity signatures coexisting within the same angular interval. Such behavior suggests that the turbulent cascade channels energy into different wave modes at different frequencies, resulting in a frequency-dependent partitioning of helicity and a scale-dependent helicity decomposition. Based on previous studies, we interpret the low-frequency negative-helicity patches as consistent with quasi-parallel, left-handed AIC waves, while the high-frequency positive-helicity features could reflect the presence of right-handed FM/W-like waves \citep{verscharen2013instabilities, bourouaine2013limits, woodham2019parallel}. To further investigate this interpretation, we next examine the ion temperature anisotropies and their potential connection to kinetic instabilities.

\subsection{Temperature Anisotropy}

\begin{figure}
	\centering
		\includegraphics[width = \columnwidth]{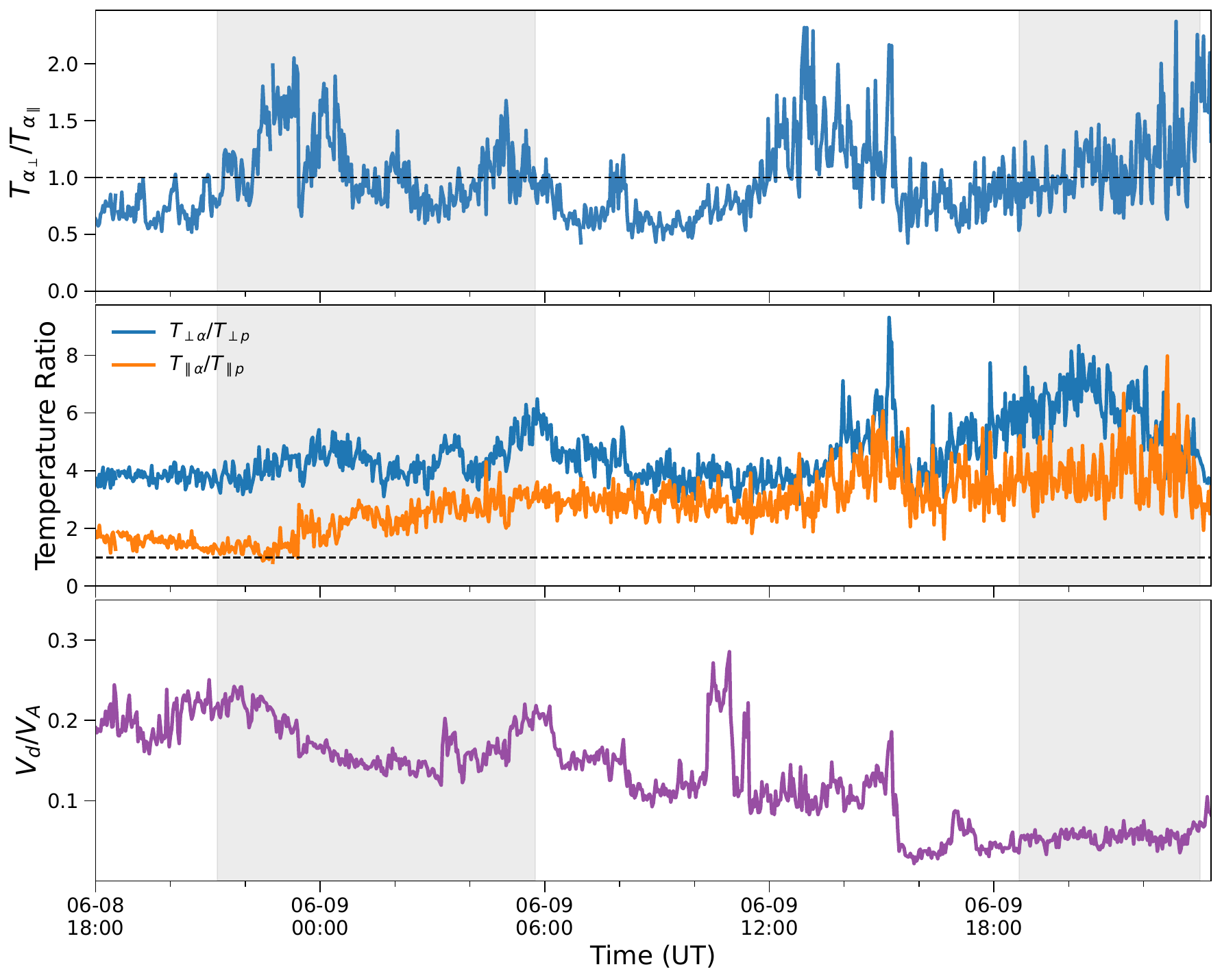}
	\caption{The upper panel of the figure shows the temperature anisotropy of alpha particle, and the middle panel depicts the ratio of perpendicular and parallel components of Alpha and proton, respectively. The bottom panel depicts ratio of  differential velocity ($V_d=V_{\alpha}-V_p$) to the Alfv\'en speed ($V_A$).}
	\label{fig:Alpha}
\end{figure}

As seen in Figure~\ref{fig:ip2}, the proton temperature anisotropy remains consistently below unity throughout the interval, indicating evidence of parallel proton heating, a feature routinely observed in the interior regions of ICMEs \citep{dasso2003parametric} and reference therein.  To assess whether preferential perpendicular heating occurs for other ion species, we analyzed the alpha-particle temperature anisotropy. The top panel of Figure~\ref{fig:Alpha} shows that $T_{\perp\alpha}/T_{\parallel\alpha}$ frequently exceeds unity in both intervals. According to \citet{verscharen2013instabilities}, alpha particles can resonate with the AIC mode when their anisotropy or the differential speed of the alpha-proton ($V_d=V_{\alpha}-V_p$) crosses the $\beta$-dependent instability thresholds.

In our case, although the alpha-particle anisotropy is greater than unity, the alpha-proton differential flow velocity is significantly smaller than the Alfvén speed (bottom panel of figure \ref{fig:Alpha}). As a result, the plasma does not exceed the $\beta$-dependent AIC thresholds of \cite{verscharen2013dispersion,verscharen2013parallel,verscharen2013instabilities}. This places both intervals in a marginally stable regime \citep{marsch1987observational,hellinger2013protons}, consistent with past or residual cyclotron-resonant signatures, but not indicative of an actively unstable AIC mode. For this reason, we examine the species-specific temperature ratios.

The middle panel shows that the perpendicular temperature ratio, \(T_{\perp\alpha}/T_{\perp p}\), exhibits enhancements within both intervals, indicating substantially stronger perpendicular heating of alpha particles compared to protons. These enhancements occur despite the very small alpha–proton differential flow, consistent with previous solar-wind observations reporting elevated alpha perpendicular temperatures at low drift speeds \citep{bourouaine2011temperature,bourouaine2013limits,chandran2013stochastic}. In contrast, the parallel temperature ratio, \(T_{\parallel\alpha}/T_{\parallel p}\), remains comparatively low especially in interval 1. This separation between perpendicular and parallel temperature ratios highlights pronounced species-dependent differences in the ion thermal properties under the observed conditions. Similar behavior has recently been confirmed in Wind observations, where alpha perpendicular heating is strongest when the differential flow is small \citep{zhao2020dependence}.

The work of \citet{bourouaine2013limits} considered the condition $V_{d} < 0.1\,V_A$ in order to minimize the contribution of drift-driven instabilities. This is consistent with our case, where the Alfv\'en speed is naturally larger than the alpha-proton differential speed (shown in last panel). Their study also showed that an enhanced parallel temperature ratio can arise when the plasma approaches the FM/W instability threshold. As seen in Figure~\ref{fig:Alpha}, during the second interval the ratio $T_{\parallel\alpha}/T_{\parallel p}$ increases unsually at the same time that $T_{\perp\alpha}/T_{\perp p}$ begins to decrease, suggesting that the elevated parallel temperature ratio may indicate the presence of FM/W-related activity.

In our case, the plasma remains marginally stable, so active AIC growth is not expected. However, the reduced magnetic helicity spectra still exhibit AIC-like quasi-parallel fluctuations near ion kinetic scales, evident in figure \ref{fig:spectral}. The ion distributions may retain signatures of earlier cyclotron-resonant heating, while the wave field reflects the presence of convected or turbulence-driven AIC-like fluctuations rather than waves generated by a local instability. Such quasi-parallel ion-scale waves are typically swept past the spacecraft from smaller heliocentric distances \citep{klein2014physical}, implying that their generation likely occurred upstream of the measurement location.

\begin{figure}
	\centering
		\includegraphics[width = \columnwidth]{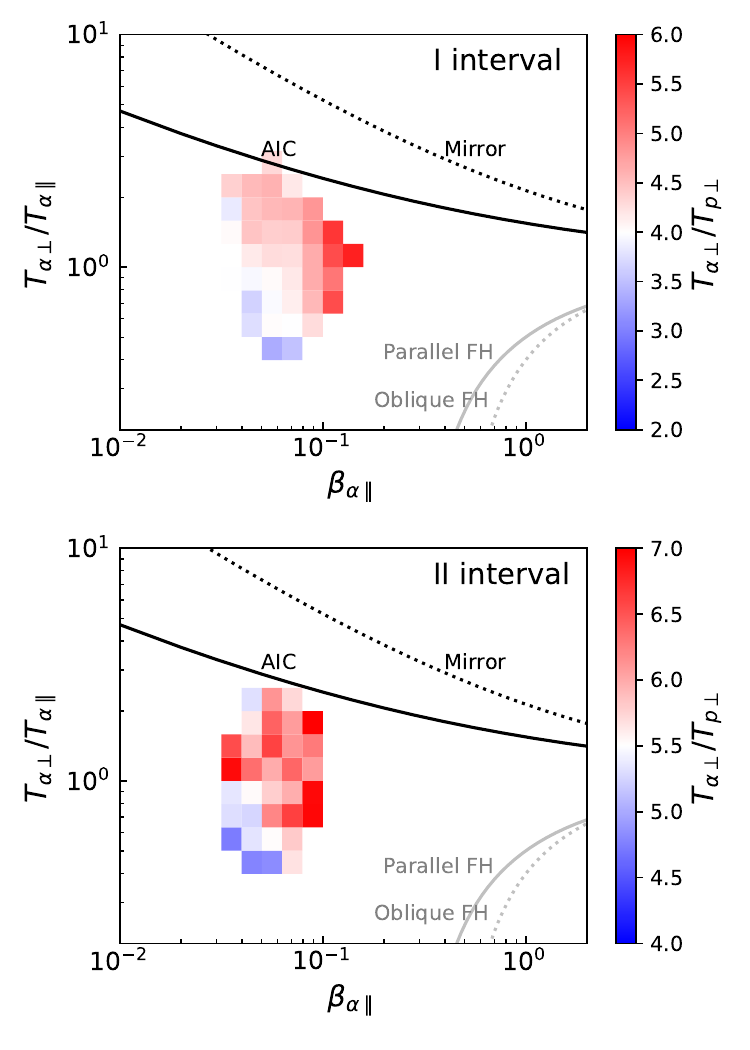}
	\caption{
Distribution of alpha-proton perpendicular temperature anisotropy $T_{\alpha\perp}/T_{p\perp}$ in the $\beta_{\alpha\parallel}$-$ T_{\alpha\perp}/T_{\alpha\parallel}$ plane for first (top panel) and second (bottom panel) time intervals. The black solid (dotted) curve indicates the AIC (mirror mode) instability thresholds. The grey solid (dotted) curve indicates the parallel (oblique) firehose instability threshold. All the thresholds are reported according to \citet{maruca2012instability}.
}

	\label{fig:instabilities}
\end{figure}

To validate our claim regarding cyclotron resonance through alpha particles, we analyzed the distribution of perpendicular temperature ratio, \(T_{\perp\alpha}/T_{\perp p}\),  in the $\frac{T_{\alpha_\perp}}{T_{\alpha_\parallel}}-\beta_{p_\parallel}$ plane.
In Figure~\ref{fig:instabilities}, it is evident that as the anisotropy of alpha particles increases, the ratio of perpendicular heating by alpha particles also rises in both intervals. Bins with counts less than 5 have been neglected in the histograms. The fitting parameters from \cite{maruca2012instability} were adopted to show the growth-rate limits of the relevant instabilities. At higher alpha-particle anisotropies, the alpha particles tend to grow in anisotropy towards the AIC instability threshold. This trend is in agreement with previous statistical results by \citet{bourouaine2013limits}, where the alpha-proton anisotropy trend has been established on a robust data analysis of about 20 years of Wind data for the low drift case, i.e., $V_d<0.1V_A$. In this framework, despite the low statistics of our ICME data samples, we clearly observe a similar trend in the $T_{\alpha\perp}/T_{\alpha\parallel}$-$\beta_{\alpha\parallel}$ plane for both interval as shown in Figure \ref{fig:instabilities}.
We then conclude that the free energy released through AIC instabilities results in preferential perpendicular heating of alpha particles rather than protons \citep{bourouaine2013limits}. These results are consistent with kinetic simulations of \cite{Valentini2016differential}, which show that turbulence naturally produces differential ion heating, with alpha particles experiencing stronger perpendicular energization than protons due to species-dependent kinetic dynamics. Regarding the parallel heating of proton, the  majority of the ICMEs have $T_{\parallel}> T_{\perp}$, reflecting expansion-driven anisotropy in the flux-rope interior rather than active instability \citep[][and reference therein]{dasso2003parametric}.

Recent studies have shown that ion-kinetic fluctuations often consist of a mixture of cyclotron-like modes and compressive FM/W modes, particularly in turbulent or inhomogeneous regions \citep{roberts2017multipoint,he2011oblique,narita2018density}. Our reduced magnetic helicity analysis reveals a similar pattern: the first interval is dominated by left-hand, quasi-parallel AIC-like signatures, whereas the second interval contains both left-hand and right-hand polarized power. Because FM/W waves are intrinsically compressive, a key diagnostic for separating them from  AIC fluctuations is the correlation between plasma density variations ($\delta n$) and magnetic-field magnitude fluctuations ($\delta |B|$) \citep{howes2012slow,roberts2017multipoint}. In particular, FM/W waves exhibit in-phase $\delta n$–$\delta |B|$ correlations, whereas kinetic Alfvén waves, though also compressive, produce $180^\circ$ phase-shifted  correlations between $\delta n$ and $\delta |B|$. This distinction has been demonstrated clearly in recent MMS observations of kinetic-scale wave activity \citep[e.g.,][]{gershman2017wave}.

\begin{figure}
	\centering
		\includegraphics[width =\columnwidth]{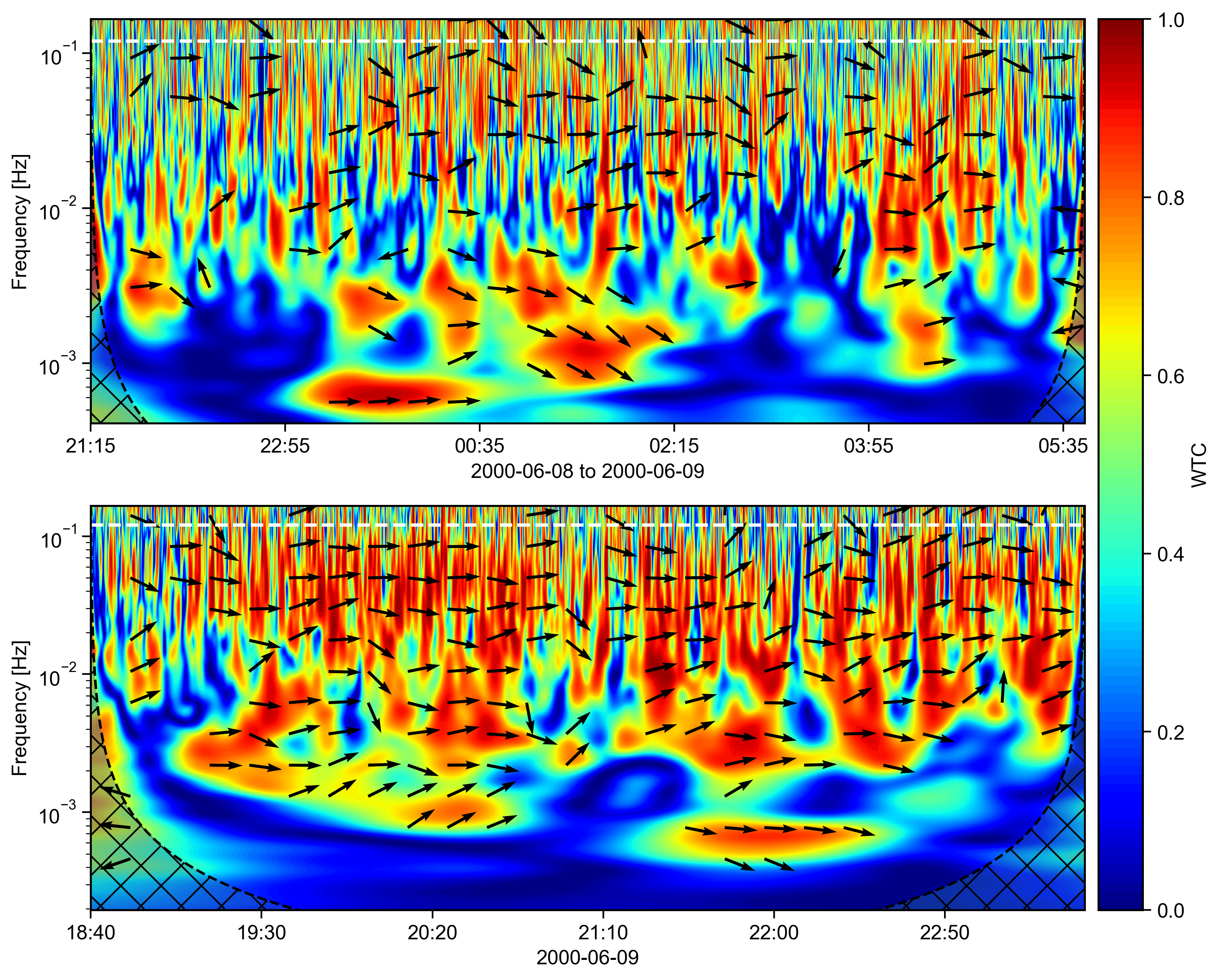}
    \caption{
    Wavelet transform coherence (WTC) between magnetic field magnitude fluctuations ($\delta |B|$) and proton density fluctuations ($\delta n$) for two studied intervals. The color scale indicates the coherence magnitude, ranging from 0 (no coherence) to 1 (perfect coherence). Black arrows denote the relative phase between $\delta |B|$ and $\delta n$: right (left) arrows correspond to in-phase (anti-phase) behavior, while upward (downward) arrows indicate that $\delta |B|$ leads (lags) $\delta n$ by $90^\circ$. The white dashed horizontal line marks the local ion cyclotron frequency ($f_{\mathrm{ci}}$). Cross-hatched regions indicate the cone of influence (COI), where edge effects may affect the reliability of the coherence estimates.}

	\label{fig:b-n}
\end{figure}

Figure~\ref{fig:b-n} presents the wavelet transform coherence (WTC) between $\delta n$ and $\delta |B|$ for the two intervals. WTC can be interpreted as the local correlation between two continuous wavelet transforms (CWTs), through which locally phase-locked behavior can be identified \citep{grinsted2004application}. The fluctuations in $n$ and $|B|$ are computed using a 10-minute rolling average. As shown in Figure~\ref{fig:b-n}, regions of high coherence (red–yellow colors) are predominantly observed at frequencies $\gtrsim 10^{-2}$~Hz, indicating strong coupling between density and magnetic compressions at higher frequencies. The black phase arrows further indicate that the fluctuations are predominantly in phase (arrows pointing to the right), consistent with a positive $\delta n$–$\delta |B|$ correlation and compressive wave behavior. The hatched regions denote the cone of influence (COI), outside of which edge effects may affect the reliability of the coherence estimates.

Both intervals exhibit coherent compressive signatures; however, the second interval displays more extended and intense high-coherence patches at high frequencies, together with a more persistent in-phase relationship. This indicates a significantly stronger compressive component in the second interval. These observations support the interpretation that the right-hand polarized population observed during the second interval could be associated with highly compressive FM/W-like fluctuations \citep{narita2018density}.

\begin{figure}
	\centering
		\includegraphics[width = \columnwidth]{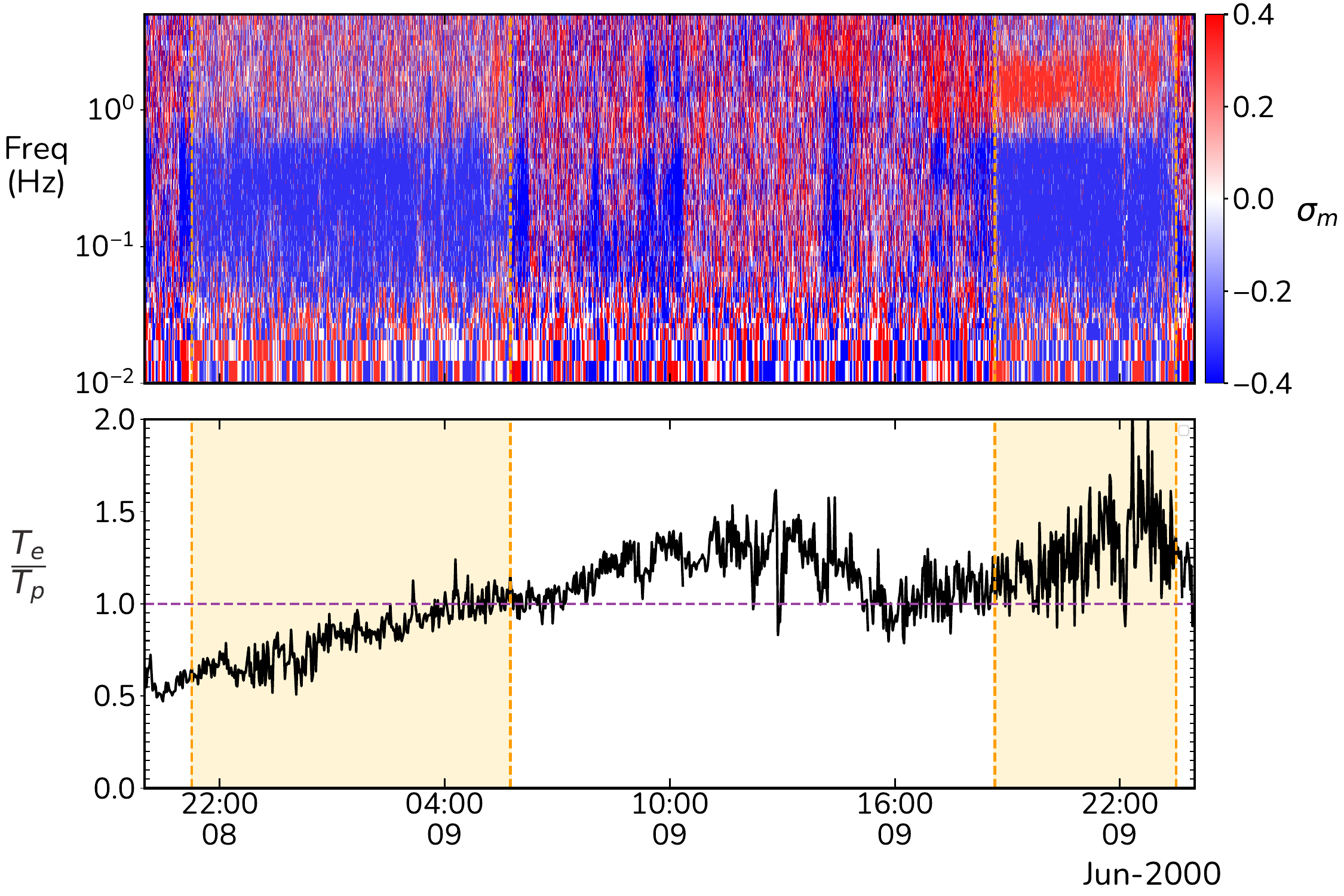}
	\caption{The upper panel of the figure shows a scalogram of the normalized magnetic helicity ($\sigma_m$). Bottom panel depicts the ratio of Electron temperature to proton temperature studied with ground moments. }
	\label{fig:scalogram}
\end{figure}

We examined various plasma parameters to identify quantitative differences between the two studied intervals. As shown in the upper panel of Figure~\ref{fig:scalogram}, the normalized magnetic helicity $\sigma_m$ was computed as a function of time and frequency. 
By examining the electron-to-proton temperature ratio we observe a transition from a region of $T_{e}/T_{p} < 1$ (first interval) to a region where the temperature of electron is dominating, i.e., $T_{e}/T_{p} > 1$ (second interval). 
This trend is depicted in the bottom panel of Figure~\ref{fig:scalogram}. Although the temperature ratio does not directly diagnose wave activity, its enhancement overlaps with the interval where mixed polarization appears. In the context of ICMEs, an increased $T_e/T_p$ ratio can modify the local pressure balance and reduce the phase speed of right–hand waves through the similar kind of “indirect” electron effect described by \citet{dasso2003parametric}, in which slower waves become more accessible to ion resonance. Such a shift would naturally lead to stronger compressive signatures, consistent with what we observe in Figure~\ref{fig:b-n}.

In compressively driven collisionless plasmas, the partition of turbulent heating between ions and electrons depends strongly on the level of compressive driving \citep{kawazura2020ion}. 
These compressive activity are closely associated with density fluctuations, which directly influence electron thermodynamics because the electron population responds more efficiently to compressive perturbations than ions \citep[e.g.,][]{schekochihin2009astrophysical,howes2012slow,chen2014intermittency}, due to higher mobility.
Recent kinetic PIC simulations further support this picture by showing that the partition of turbulent energy between electrons and ions depends on their relative temperatures \citep{adhikari2025kinetic}. 

     \section{Discussion}

This work presents new evidence that, during the analyzed periods, ICME fluctuations exhibit localized turbulent features, departing from various characteristics that are generally associated with magnetic clouds.
Whether this structure forms in the lower corona or develops during propagation, with the same or different source of origin remains unclear. Several characteristics, including normalized cross helicity (\( \sigma_c \)), normalized residual energy (\( \sigma_r \)), power spectral densities, and drift velocity (some are not shown here), exhibit similar behavior; however, the normalized magnetic helicity behaves differently and, in the second interval, reveals a mixed population. This motivates us to explore what determines the preferred dissipation pathways in CME-driven turbulence and how different wave modes contribute to energy transfer at kinetic scales. 

Fluid-scale Alfvénic fluctuations in  
plasma are key in driving kinetic-scale wave activity. 
Alfvénic turbulence is inherently anisotropic, with its characteristics significantly influenced by the angle $\theta_{VB}$. Turbulent dissipation is expected to produce plasma heating that reflects the spatial distribution of fluctuations at ion-kinetic scales. The polarization properties of these fluctuations determine the nature of dissipation mechanisms. 
High-frequency Alfvén waves dissipate resonantly, transferring energy to ions and preferentially heating them in the direction perpendicular to the magnetic field, which in turn enhances their temperature anisotropy \citep{telloni2019ion}. The amplitude of Alfvén waves saturates as the thermal anisotropy approaches equilibrium \citep{zhao2021mhd,vech2021wave}. Consequently, the  velocity distribution can cross the cyclotron instability threshold, triggering the  generation of AIC waves and facilitating the release of excess perpendicular thermal energy \citep{telloni2019ion,telloni2020wave}. In the present study, however, we find that the observed AIC waves are more likely driven by alpha-particle temperature anisotropy rather than by protons 
\citep[e.g.,][]{bourouaine2011temperature,bourouaine2013limits}.



The FM/W instability is an ion–beam–driven instability that arises when drifting proton beams resonantly interact with right-hand circularly polarized electromagnetic waves propagating parallel to the background magnetic field \citep{daughton1998electromagnetic}. Through the cyclotron-resonance condition, the beam transfers energy to the wave field, producing wave growth near the proton cyclotron frequency. The instability becomes efficient once the beam velocity exceeds a threshold fraction of the Alfvén speed \citep{gurnett2005introduction}. Consequently, FM/W modes are most easily excited in high-beta conditions or whenever the proton beam provides sufficient free energy for right-hand polarized fluctuations. In multi–ion plasmas, sometimes both temperature anisotropy and differential flow supply free energy not only to the left–hand AIC branch but also to the right–hand FM/W branch. When $T_{\perp}/T_{\parallel}<1$ or when even a modest ion beam is present, the FM/W mode is preferentially destabilized, enabling the beam or anisotropic population to transfer kinetic energy into electromagnetic fluctuations  \citep{li2000proton}. 
Recent kinetic simulations by \citet{khoosheshahi2025thermal} demonstrate that proton beams can directly excite parallel FM/W waves, with thermal effects increasing their phase speed. Similarly, \citet{scholer2007whistler} shows that differential flow between core and beam ions drives FM/W growth through cyclotron resonance, resulting in beam deceleration and scattering. Therefore, the right hand polarized signatures observed in the second interval, together with enhanced compressibility, are consistent with anisotropy regulated FM/W activity. 

The coexistence of left-hand and right-hand polarized signatures does not always require active local instabilities. As shown by \citet{klein2014physical}, the polarization and phase relationships of ion-scale fluctuations are primarily governed by the linear eigenfunctions of kinetic modes, even when the plasma is marginally stable. Under this quasilinear framework, the turbulent cascade can naturally populate AIC-like and FM/W-like branches, allowing linear-mode helicity signatures to appear in the absence of local wave growth.
{\citet{woodham2019parallel} showed that proton-kinetic-scale magnetic helicity can be used to distinguish fluctuations compatible with AIC and FM/W modes from those generated by the oblique turbulent cascade. Given the parallel heating of proton  in our intervals, the observed ion-scale signatures are more consistent with fluctuations influenced by the alpha population. Although the anistropy is
found empirically to be weak when the  ion differential speed is small, the alpha particles can still be perpendicularly
heated much more than the protons \citep{bourouaine2011temperature}.  Kinetic simulations support this interpretation: hybrid models \citep{maneva2015relative} demonstrate that turbulent Alfvén–cyclotron fields preferentially heat alpha particles, even when differential flow is negligible, while Vlasov–Maxwell simulations \citep{Valentini2016differential} show that coherent structures efficiently generate species-dependent ion energization at kinetic scales. Although Ulysses analyses by \citet{matteini2013signatures} focus on proton core–beam regulation, they indicate that ion distributions frequently evolve toward marginal stability while retaining measurable nonthermal features. Such residual nonthermal features implies the presence of free energy that can be available to drive kinetic-scale activity. In this context, the alpha-particle properties observed here are consistent with a plasma that has undergone earlier perpendicular energization, with anisotropic alpha populations supplying the source of free energy for the ion-scale fluctuations identified in our ICME intervals. Although the exact origin of the enhanced alpha-particle temperatures in these ICME intervals is uncertain, cyclotron-heating and stochastic-heating models can account for the preferential heating of alpha particles to temperatures exceeding those of protons.

Previous studies have shown that the ICME sheath region, is often populated with AIC waves \citep{farris1993magnetic}. In contrast, upstream regions generally exhibit FM/W modes. Similar Alfvénic fluctuations have also been observed in the sheath and inside CMEs \citep{dhamane2024situ,farrugia2020study}. The presence of kinetic-scale waves has a profound influence on the internal energy and thermodynamics of the coherent, large-scale structure of ICME magnetic clouds. AIC waves resonate with ions at their cyclotron frequency, driving ion heating and temperature anisotropies that trigger micro-instabilities and intensify turbulence. Meanwhile, FM/W waves transfer energy efficiently from electromagnetic modes to particles, causing localized heating and anisotropic pressure distributions. 
The enhanced turbulence activity accelerates magnetic reconnection, releasing stored magnetic energy as kinetic and thermal energy, thereby reshaping the magnetic topology and energizing particles.

\section{Acknowledgments}
We acknowledge the use of the CDAWeb or ftp service by the Space Physics Data Facility of NASA/GSFC. S.B. and D.T. acknowledge the Space It Up! project funded by the Italian Space Agency and the Ministry of University and Research under contract n. 2024-5-E.0—CUP n. I53D24000060005. A.S. is partially supported by NASA FINESST award 80NSSC24K1867. O.P. acknowledges the support of the PRIN 2022 project ``The ULtimate fate of TuRbulence from space to laboratory plAsmas (ULTRA)'' (2022KL38BK, Master CUP: B53D23004850006) by the Italian Ministry of University and Research, funded under the National Recovery and Resilience Plan (NRRP), Mission 4 – Component C2 – Investment 1.1, ``Fondo per il Programma Nazionale di Ricerca e Progetti di Rilevante Interesse Nazionale (PRIN~2022)'' (PE9) by the European Union – NextGenerationEU.

\bibliographystyle{apalike}
\bibliography{ref}



\end{document}